\documentclass[prb,twocolumn,showpacs,amsmath,amssymb]{revtex4}

\usepackage{graphicx}
\usepackage{dcolumn}
\usepackage{bm}
\usepackage{epsf}
\usepackage{color}
\usepackage{amsmath}
\usepackage{eufrak}
\usepackage{mathrsfs}
\usepackage{amssymb}
\usepackage{subfigure}

\newcommand{\bcen}{\begin{center}}
\newcommand{\ecen}{\end{center}}
\newcommand{\btab}{\begin{tabular}}
\newcommand{\etab}{\end{tabular}}
\newcommand{\bdes}{\begin{description}}
\newcommand{\edes}{\end{description}}

\newcommand{\beq}{\begin{equation}}
\newcommand{\eeq}{\end{equation}}
\newcommand{\bea}{\begin{eqnarray}}
\newcommand{\eea}{\end{eqnarray}}
\newcommand{\non}{\nonumber}

\newcommand{\bary}{\begin{array}}
\newcommand{\eary}{\end{array}}
\newcommand{\benum}{\begin{enumerate}}
\newcommand{\eenum}{\end{enumerate}}
\newcommand{\bitem}{\begin{itemize}}
\newcommand{\eitem}{\end{itemize}}
\newcommand{\btau}{\mbox{\boldmath $ \tau $}}

\newcommand{\D}[1]{\mbox{d}{#1}}

\newcommand{\mean}[1]{\langle #1 \rangle}

\newcommand{\prn}[1] {(\ref{#1})}
\newcommand{\sect}[1] {Section~\ref{#1}}

\newcommand{\fig}[1]{fig.~\ref{#1}}
\newcommand{\Fig}[1]{Fig.~\ref{#1}}
\newcommand{\sG}{\mathscr{G}}
\newcommand{\cG}{{\cal G}}
\newcommand{\mG}{\mathbb{G}}
\newcommand{\mOne}{\mbox{\boldmath $1$}}

\newcommand{\bSig}{\mbox{\boldmath $ \Sigma $}}
\newcommand{\bDelta}{\mbox{\boldmath $ \Delta $}}

\newcommand{\figsmallwidth}{8.5truecm}

\newcommand{\citeasnoun}[1]{\citeauthor{#1}\cite{#1}}

\begin{document}

\preprint{}

\title{Phase Diagram of the Attractive Hubbard Model with Inhomogeneous Interactions}

\author{Vijay B. Shenoy}\email{shenoy@physics.iisc.ernet.in}
\affiliation{Centre for Condensed Matter Theory, Department of
  Physics, Indian Institute of Science, Bangalore 560 012, India}

\date{\today}

\begin{abstract}
The phase diagram of the attractive Hubbard model with spatially inhomogeneous interactions is obtained using a single site dynamical mean field theory like approach. The model is characterized by three parameters: the interaction strength, the active fraction (fraction of sites with the attractive interaction), and electron filling. The calculations indicate that in a parameter regime with intermediate values of interaction strength (compared to the bare bandwidth of the electrons), and intermediate values of the active fraction, ``non-BCS'' superconductivity is obtained. The results of this work are likely to be relevant to many systems with spatially inhomogeneous superconductivity such as strongly correlated oxides, systems with negative $U$ centers, and, in future, cold atom optical lattices.
\end{abstract}


\maketitle

\section{Introduction}

There is a growing body of experimental evidence that strongly
correlated oxides such as cuprates,\cite{Sigmund1994,Howald2001,Pan2001,Hanaguri2004,Vershnin2004,McElroy2005,McElroy2005a,Gomes2007,Gomes2007a} 
manganites\cite{Dagotto2003,Mathur2003, Dagotto2005a, Shenoy2006, Shenoy2007, Shenoy2008} etc., are electronically inhomogeneous. The term ``electronically inhomogeneous'' is used to describe states with spatially
inhomogeneous electronic orders, i.e., of orders of charge, spin,
superconducting gap etc. There are suggestions that such inhomogeneous
electronic states are one of the {\it characteristic features}
intrinsic to strongly correlated materials\cite{Dagotto2005} arising
out of their ``electronic softness''.\cite{Milward2005} A clear
understanding of this phenomenon could, for example, suggest possibilities of
controlling the nature and size of the electronic inhomogeneities, and can
lead to, inter alia, possible device applications of these materials. Efforts directed towards
uncovering the physics of the origin and nature of electronically
inhomogeneous electronic states, therefore, have emerged as a very active research area.

Of particular interest to this work is electronically inhomogeneous
superconducting state which is found in many systems of current
interest. High temperature superconducting cuprates are prominent
examples of systems showing inhomogeneous superconductivity. The past
five or so years have witnessed  fascinating  experimental work based
on scanning probes that have revealed a wealth of information
regarding the nature of the inhomogeneous superconducting state in
cuprates.\cite{Howald2001,Pan2001,Hanaguri2004,Vershnin2004,McElroy2005,McElroy2005a,Gomes2007,Gomes2007a}
In particular the experimental work reported in references
 [\onlinecite{Gomes2007,Gomes2007a}] has clearly demonstrated a
distribution of gaps, and even regions with gaps above the
superconducting transition temperatures. There could be several
physical origins to this phonemenon, such as one body disorder due to
the dopant ions, inhomogeneous pairing interactions etc.
Superconductivity arising out of inhomogeneous pairing interactions
are also found in many other systems. \citeasnoun{Anderson1975}
suggested the possibility of negative-$U$ centres in
semiconductors. There are reports of existence of superconductivity in
silicon based nanostructures with negative-$U$
centres.\cite{Bagarev2006} Negative-$U$ models have been used to describe the physics of doped bismuthates.\cite{Taraphder1995} A material of more recent interest, Tl-doped
PbTe, is believed to have a distribution of negative-$U$
centers.\cite{Matsushita2005,Matsushita2006}

\begin{figure}
\subfigure[]{\epsfxsize=\figsmallwidth \epsfbox{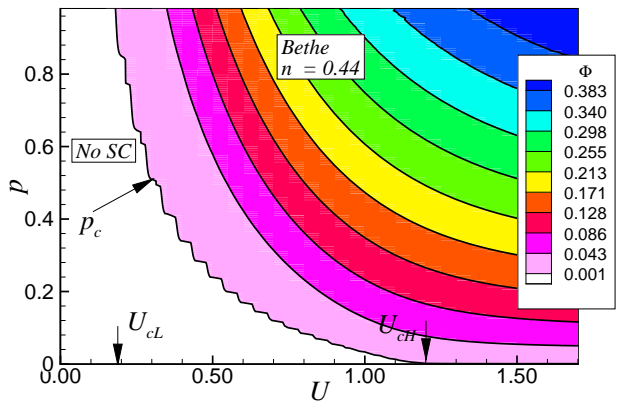}}
\subfigure[]{\epsfxsize=\figsmallwidth \epsfbox{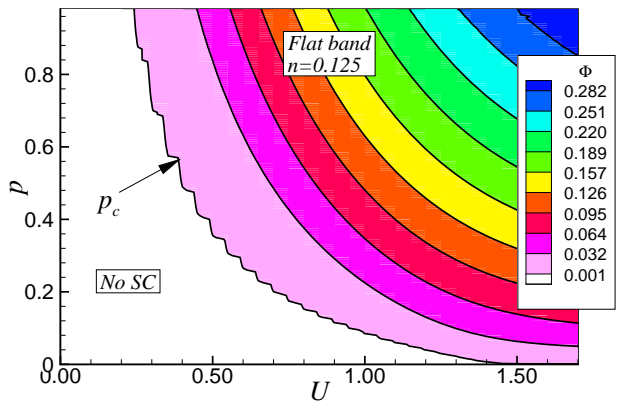}}
\caption{(color online) $T=0$ phase diagram in the $p-U$ plane. Plots show contours of constant pairing  $\Phi$. The pairing amplitude vanishes on the ``left'' side of the curve marked as $p_c$. (a) Bethe lattice with filling  $n=0.44$ (b) Flat band with filling $n=0.125$. }
\label{pUTzero}
\end{figure}

As indicated briefly above, there are two factors that lead to an
inhomogeneous superconducting state. The first one is one body
disorder; when one body disorder is large, it tends to localize
electrons, and in this sense ``competes'' with
superconductivity.\cite{Ma1985} Effect of one body disorder on
superconductivity has been extensively studied using models and
methods of different
sophistication.\cite{Ghosal2001,Valdez-Balderas2006,Garg2006} 
The second factor that contributes to inhomogeneous superconductivity arises in situations where the pairing interaction (such as negative-$U$ centers) responsible for superconductivity is itself spatially inhomogeneous.
 Systems
with a distribution of negative-$U$ centers are known to give rise to the ``charge Kondo
effect''\cite{Taraphder1991}; the material Tl-doped PbTe is believed
to be one such.\cite{Dzero2005}
Models
with inhomogeneous pairing interactions have been investigated before.\cite{Rice1981,vanderMarel1992,Litak2000,Nunner2005,Andersen2006,Aryanpour2006,Aryanpour2007,Chatterjee2007} The model usually studied is the attractive Hubbard model with inhomogeneous interactions (AHII) is described by the Hamiltonian:
\bea
H = -t \sum_{<ij> \sigma} \left( c^\dagger_{i\sigma} c_{j \sigma} + \mbox{h.c.} \right)  - \sum_{i} U_i n_{i \uparrow} n_{i \downarrow} - \mu \sum_{i \sigma} n_{i \sigma} \label{rhub}
\eea
where $i,j$ are site indices of a lattice, $t$ is the hopping
amplitude, $\sigma$ is the spin index, $c^\dagger_{i \sigma}$ is the
electron operator which creates an electron of spin $\sigma$ at site
$i$, $n_{i \sigma} = c^\dagger_{i \sigma} c_{i\sigma}$ is the number
operator at site $i$ of spin $\sigma$, $\mu$ is the chemical
potential. The interaction $U_i \ge 0$ is site dependent; a fraction
$p$ (here called the active fraction) of sites have $U_i = U >0$, while $U_i =0$ for the other fraction
$(1-p)$ of sites. These sites with $U _i \ne 0$ can be arranged
periodically or randomly. It is known that for a given $U$, there is a
critical value $p_c$ of $p$ below which superconductivity is
killed.\cite{Rice1981,vanderMarel1992,Litak2000} 

More recently the above model \prn{rhub}, motivated by the electronic inhomogeneities in correlated materials,  has been subjected to
extensive numerical
simulations.\cite{Aryanpour2006,Aryanpour2007,Chatterjee2007} A
Bogoliubov-de Gennes mean-field (BdGMF) approach is used to obtain the
ground state and finite temperature properties; these calculations
involve averaging over several different $U$-disorder
realizations. Quantum Monte Carlo calculations\cite{Hurt2005} have also been performed on a two-dimensional square lattice. Such calculations are numerically intensive, and
attention has been focussed on particular values of electron fillings
and interaction parameters $U$, and the active fraction $p$. It is
useful to have a ``phase diagram'' of the AHII, particularly to compare and
contrast different experimental systems, and to obtain regions in the
parameter space where interesting physics may be expected. Calculation
of the phase diagram within the BdGMF approach can be quite time
consuming; it is therefore desirable to generate the phase diagram by means of a simple approach
to understand its overall structure.

\begin{figure}
\subfigure[]{\epsfxsize=\figsmallwidth \epsfbox{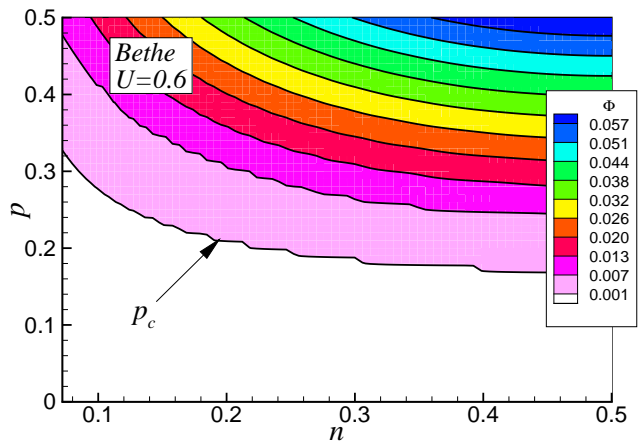}}
\subfigure[]{\epsfxsize=\figsmallwidth \epsfbox{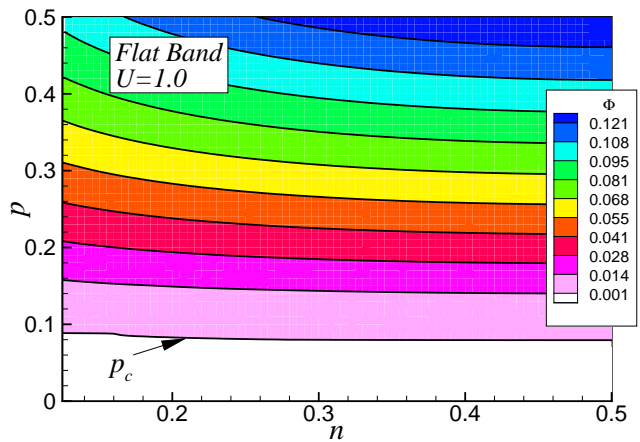}}
\caption{(color online) $T=0$ phase diagram in the $p-\mu$ plane. Plots show contours of constant pairing  $\Phi$. The pairing amplitude vanishes below the curve marked as $p_c$. (a) Bethe lattice with $U=0.6$  (b) Flat band with $U=1.0$.
}
\label{pmuTzero}
\end{figure}

Motivated by the above discussion, the phase diagram of the AHII model
is obtained in this paper using a dynamical mean field theory like
approach.\cite{Georges1996} Effects of interaction strength, active
fraction and electron filling are investigated systematically. It is
found that there are regions of the phase diagram where
``unconventional'' superconductivity is supported.

The paper is organized as follows. In the next section
(\sect{formulation}), we discuss the single site formulation of the
problem. Results of the calculations are presented in
\sect{Results}. In the final section (\sect{discussion}) the results
are discussed, and aspects of the phase diagram not captured by the
present treatment, including a more speculative phase diagram, are
discussed.

\section{Formulation}
\label{formulation}

The AHII Hamiltonian given in \prn{rhub} is treated here within a
dynamical mean field theory like approach. A dynamical mean field theory treatment of Hubbard like Hamiltonians with {\it one body} disorder is treated by \citeasnoun{Janis1992}; the present work deals with the case where the {\it interaction} term is disordered. A single site, that
hybridizes with an electron bath\cite{Georges1996} described by a bath
Green's function $\cG$ (this is a matrix in the present formulation,
see below), represents a typical site of the lattice. The imaginary
time action for this site is written as
\bea
S[\Psi^\star,\Psi,V] & = &  -\int_0^\beta \int_0^\beta\, \D{\tau} \,\D{\tau'} \,  \Psi^\star(\tau) \cG^{-1}(\tau - \tau')  \Psi(\tau') \non \\
& & - \int_0^\beta \,\D{\tau} \, V \psi^\star_{\uparrow}(\tau) \psi^\star_{\downarrow}(\tau) \psi_{\downarrow}(\tau) \psi_{\uparrow}(\tau)
\eea
where $\beta = 1/T$ ($T$ is the temperature), $\psi_\sigma$ are the
Grassmann variables of the site electrons,
\bea
\Psi = \left( \begin{array}{c} \psi_\uparrow \\ \psi^\star_\downarrow \end{array} \right), \,\, \Psi^\star = \left( \begin{array}{c} \psi^\star_\uparrow \,\,\, \psi_\downarrow \end{array} \right) \non
\eea
are the Nambu matrices, $\cG$ is the matrix Green's function (which
incorporates the chemical potential $\mu$), $V$ is the attractive
interaction. The Grassmann variables satisfy the Fermionic condition
$\psi_\sigma(\beta) = -\psi_\sigma(0)$. The interaction $V$ is a
random variable which here is distributed according to the probability
distribution
\bea
P(V) = (1-p) \, \delta(V) + p \,\delta(V -U)
\eea
where $p$ is the active fraction,
$\delta(\cdot)$ is the Dirac delta function, and $U$ is the attractive
interaction strength at the active sites. The disorder averaged
partition function can now be written as a disorder averaged path
integral
\bea
Z = \int \D{V} \, P(V) \, \int {\cal D}[\psi^\star_\sigma,\psi_\sigma] \,e^{-S[\Psi^\star,\Psi,V]}
\eea
The site Green's function (expressed in terms of Matsubara frequencies $i \omega_n$) is obtained as
\bea
\mG(i \omega_n)  = (1-p) \, \cG(i \omega_n) + p \, (\cG^{-1}(i \omega_n) - \Sigma_U(i \omega_n))^{-1} \label{siteGF}
\eea
where $\Sigma_U$ is the self energy obtained from the solution of the
quantum impurity problem with the bath Green's function $\cG$ and an
attractive Hubbard interaction $U$ at the site. The Green's function \prn{siteGF} represents the lattice Green's function as seen from the single site formulation. The site self energy
is now given by
\bea
\bSig(i \omega_n) = \cG^{-1}(i\omega_n) - \mG^{-1}(i \omega_n). \label{siteSelf}
\eea

Using the dynamical mean field theory ansatz\cite{Georges1996} that the
the self energy is ``momentum (energy) independent'', we obtain the
lattice Green's function $\sG$ as
\bea
\sG(i\omega_n) = \int \, \D{\varepsilon} \, g(\varepsilon) \, \left(i \omega_n \, \mOne - \xi \, \btau_z -\bSig(i \omega_n) \right)^{-1} \label{Hiltrans}
\eea
where  $\mOne$ is  a $2\times2$  unit  matrix, $\btau_z$  is the  Pauli
$z$-matrix,  $\xi = \varepsilon  - \mu$,  and $g(\varepsilon)$  is the
bare density of states of the lattice. The dynamical mean field theory
self consistency condition is now obtained by insisting that
\bea
\cG^{-1}(i\omega_n) = \sG^{-1}(i\omega_n) + \bSig(i\omega_n). \label{selfCons}
\eea
This condition is obtained by demanding that the site Green's function as calculated from the quantum impurity formulation \prn{siteGF} is same as the lattice Green's function calculated via \prn{Hiltrans}.

The following paragraphs contain a description of the approximate
solution, based on the saddle point method, of the quantum impurity
problem that is used in this work. A Hubbard-Stratanovich field
$\Delta(\tau)$ is introduced to decouple the interaction term in the
particle-particle channel. The partition function becomes
\begin{widetext}
\bea
Z =  \int \D{V} \, P(V) \, \int {\cal D}[\Delta^*,\Delta] \, e^{- \int_0^\beta \, \D{\tau} |\Delta(\tau)|^2} \int {\cal D}[\psi^\star_\sigma,\psi_\sigma] \, e^{-S[\Psi^\star,\Psi,V,\Delta]}
\eea
where
\bea
S[\Psi^\star,\Psi,V,\Delta] = - \int_0^\beta \int_0^\beta \D{\tau} \D{\tau'} \, \Psi^\star(\tau) \, \left(\cG^{-1}(\tau - \tau')  + \sqrt{V} \bDelta(\tau - \tau') \right) \, \Psi(\tau')
\eea
with $\bDelta$ defined as
\bea
\bDelta(\tau - \tau') = \left( \begin{array}{cc} 0 & \Delta(\tau) \\ \Delta^*(\tau) & 0 \end{array} \right) \delta(\tau - \tau'). \label{Deldefn}
\eea
The Fermionic path integral is easily evaluated, and the partition function becomes
\bea
 \int \D{V} \, P(V) \, \int {\cal D}[\Delta^*,\Delta] \, e^{- \left(\int_0^\beta \, \D{\tau} |\Delta(\tau)|^2 - \ln \det\left[ -\left( \cG^{-1} + \sqrt{V} \bDelta \right) \right] \right)}.
\eea
The averaging  over the probability distribution can be performed {\it exactly} to obtain
\bea
Z = \int {\cal D}[\Delta^*,\Delta] e^{-S_\Delta}
\eea
where
\bea
S_\Delta = \int_0^\beta \, \D{\tau} |\Delta(\tau)|^2 - \ln\left[ (1 -p) \det(-\cG^{-1}) + p \det(-(\cG^{-1} + \sqrt{U} \bDelta )) \right].
\eea

The partition function is now evaluated by introduction of the saddle
point approximation which amounts to treating $\Delta$ as independent
of the imaginary time\cite{Hubbard1959}; in the present context this approximation is equivalent to the BCS mean-field decoupling of the interaction term. With the assumption that $\Delta$ is real
(equivalent to picking a particular phase of the resulting
superconductor), the value of $\Delta$ is obtained by minimizing
$S_\Delta$ leading to the equation
\bea
\Delta = \frac{p e^{\ln \det[-(\cG^{-1} + \sqrt{U} \bDelta )]}}{ (1 - p) e^{\ln \det[-\cG^{-1}]} + p  e^{\ln \det[-(\cG^{-1}  + \sqrt{U} \bDelta )]}} \left[\frac{\sqrt{U}}{2} \frac{1}{\beta} \sum_{i\omega_n} \left(\cG^\Delta_{12}(i \omega_n) + \cG^\Delta_{21}(i \omega_n) \right) \right] \label{DelSPEqn}
\eea
where $\cG^\Delta = (\cG^{-1} + \sqrt{U} \bDelta)^{-1})$.
\end{widetext}
Within this approximation the self energy $\Sigma_U$ in \prn{siteGF}
is equal to $\sqrt{U} \bDelta$, where the $\Delta$ obtained from the
solution of \prn{DelSPEqn} is used in \prn{Deldefn}.

In present formulation within a dynamical mean field theory framework,
the saddle point approximation is the simplest possible ``impurity
solver''. The formulation based on the Hubbard-Stratanovich fields is
amenable to more sophisticated, and obviously more computationally
intensive, treatments such as the Hrisch-Fye quantum monte carlo
method.\cite{Hirsch1986} The saddle point approximation for the
impurity solver is similar to the coherent potential
approximation.\cite{Elliott1974}

In the framework developed here, $\sqrt{U} \Delta$ has the natural
interpretation of the disorder averaged pairing gap. It should
also be noted that there is a possibility of introducing a second
Hubbard-Stratanovich field in the particle-hole channel, which in
effect is equivalent to introducing an additional Hartree potential in
the saddle point approximation. This extra Hartree potential can now
be absorbed into the definition of the chemical potential $\mu$.

For a given value of $U$ and $p$, the value of $\Delta$ is calculated
as follows. A typical calculation starts with an assumed value of
$\Delta$, and a new value of $\Delta$ is calculated using
\prn{DelSPEqn}. The site Green's function \prn{siteGF} and the self
energy \prn{siteSelf} are calculated using the new value of
$\Delta$. The site self energy is used in \prn{Hiltrans} to obtain the
lattice Green's function, and a new bath Green's function is generated
using \prn{selfCons}. This process is carried out until the values of
$\Delta$, self energy $\bSig$ are within a specified tolerance of each
other in two successive iterations. All self-consistency calculations
are done at fixed chemical potential $\mu$ and the number of electrons
$n$ is obtained after convergence is obtained. The chemical potential
is then adjusted to that the number electrons is obtained to be the
desired value.  The calculations reported here are performed by
evaluating all Matsubara sums as integrals along the real frequency
axis, and the self consistency condition also enforced on the real
frequency axis.

\section{Results}
\label{Results}

In the single site formulation presented in the last section, the
information regarding the lattice enters the formulation only via the
density of states $g(\varepsilon)$.  Since the goal of this paper is to understand
the generic features of the phase diagram of the AHII model, 
densities of states with simple analytical forms  that  capture some
features of the real lattice systems are adopted. Two cases are considered. The semicircular density of states corresponding to a Beth\'e lattice\cite{Mahan2000} with
\bea
g(\varepsilon) = \frac{2}{\pi} \sqrt{1 - \varepsilon^2}, \;\;\;\; -1 \le \varepsilon \le 1
\eea 
and the flat band density of states
\bea
g(\varepsilon) = \frac{1}{2}, \;\;\;\; -1 \le \varepsilon \le 1.
\eea
Energy is measured in the units of half bandwidth of the systems, and
hence the condition $-1 \le \varepsilon \le 1$ in both
cases. Other parameter values of quantities such as $U$, $\mu$,
$T$ are all henceforth dimensionless ratios of these quantities and the half bandwidth.

Superconductivity is monitored by computing the pairing amplitude $\Phi$:
\bea
\Phi = \mean{\psi^{\star}_{\uparrow} \psi^{\star}_{\downarrow}} = - \frac{1}{\pi} \int_{-\infty}^{\infty} \D{\omega} \, \Im{\mG_{12}(\omega)}
\eea
Vanishing of $\Phi$ implies absence of pairing and
superconductivity. Clearly, existence of a nonzero value of $\Phi$
automatically does not imply global superconductivity. This point is
discussed in more detail later in the paper when the Bose-Einstein
condensation like phenomenon in such systems is discussed.

Results at zero temperature are presented first followed by results
at $T > 0$.

\begin{figure}
\centerline{\epsfxsize=\figsmallwidth \epsfbox{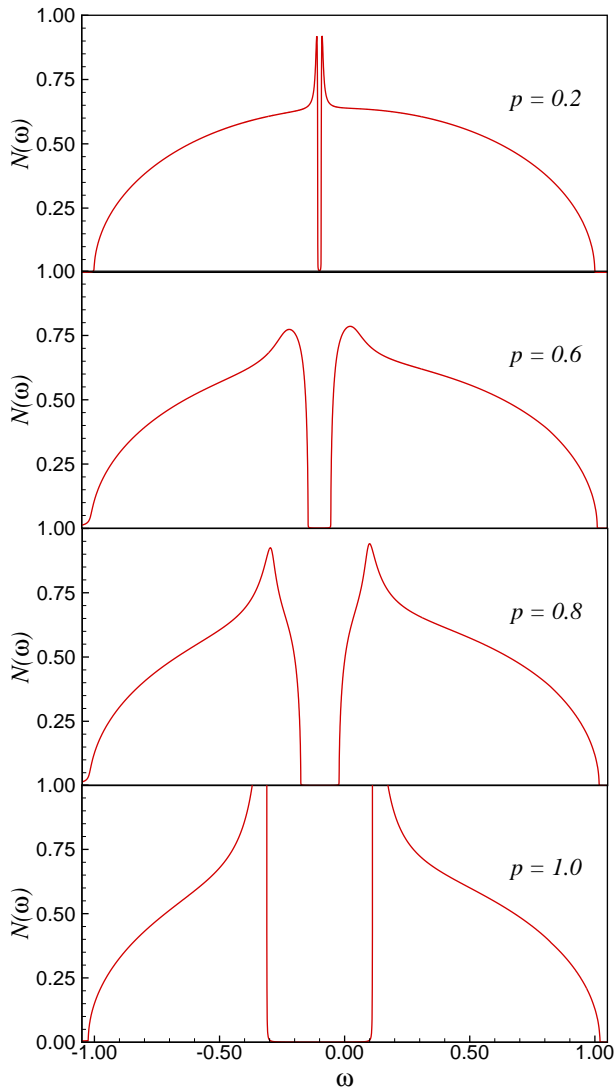}}
\caption{ (color online) Spectral function for the Bethe lattice with $U=0.8$, $n=0.44$. The four panels show the spectral function for increasing values of the active fraction $p$.}
\label{bspecwt}
\end{figure}

\subsection{Results: $T=0$}

 For a given density of states, the phase diagram is
determined by three parameters: the interaction strength $U$, the
active fraction $p$ and the filling $n$.
\Fig{pUTzero} shows the phase diagram of the system in the $p$--$U$
plane for different fillings. For a given filling $n$ we see
that there is a range of interaction strength $U$ for which there is a
critical value of the active fraction $p_c$ that is required to
produce a nonzero pairing amplitude $\Phi$. In the case of the Bethe
lattice the value of $p_c$ decreases with increasing filling $n$, while for the flat band case,
$p_c$ is essentially insensitive to filling. This result is a reflection
of the fact that $p_c$ is affected by the bare density of states (at
the chemical potential); the bare density of states increases with
increase in filling (up to $n \le 0.5$) for the Bethe lattice, and
hence the decrease of $p_c$. The critical active fraction $p_c$ is
insensitive to filling in the flat band case since the density of
states is constant. This is more clearly illustrated in \Fig{pmuTzero}
which shows the phase diagram in the $p-n$ plane for two values of
$U$. It is evident that $p_c$ increases with decreasing $n$, and this
increase is related to the decrease in the bare density of states (see
\Fig{pmuTzero}(a)). Interestingly, the sensitivity of $p_c$ on $n$
decreases with increasing interaction strength $U$. The present
calculation reveals an interesting new result. For values of $U$
larger than a critical value $U_{cH}$ (this value depends on the
filling, i.e., bare density of state at the chemical potential), the
critical value of the active fraction $p_c$ becomes vanishingly small
(see \Fig{pUTzero}). Thus if $U > U_{cH}$, even a small concentration
of impurities can produce a non-zero pairing amplitude. In the same
vein, there is another critical value of the interaction $U_{cL}$. If
the interaction strength is below $U_{cL}$ (which, again, depends on
the bare density of states at the chemical potential) even a small
dilution of the active fraction from unity kills the pairing
amplitude!

\begin{figure*}
\subfigure[]{\epsfxsize=\figsmallwidth \epsfbox{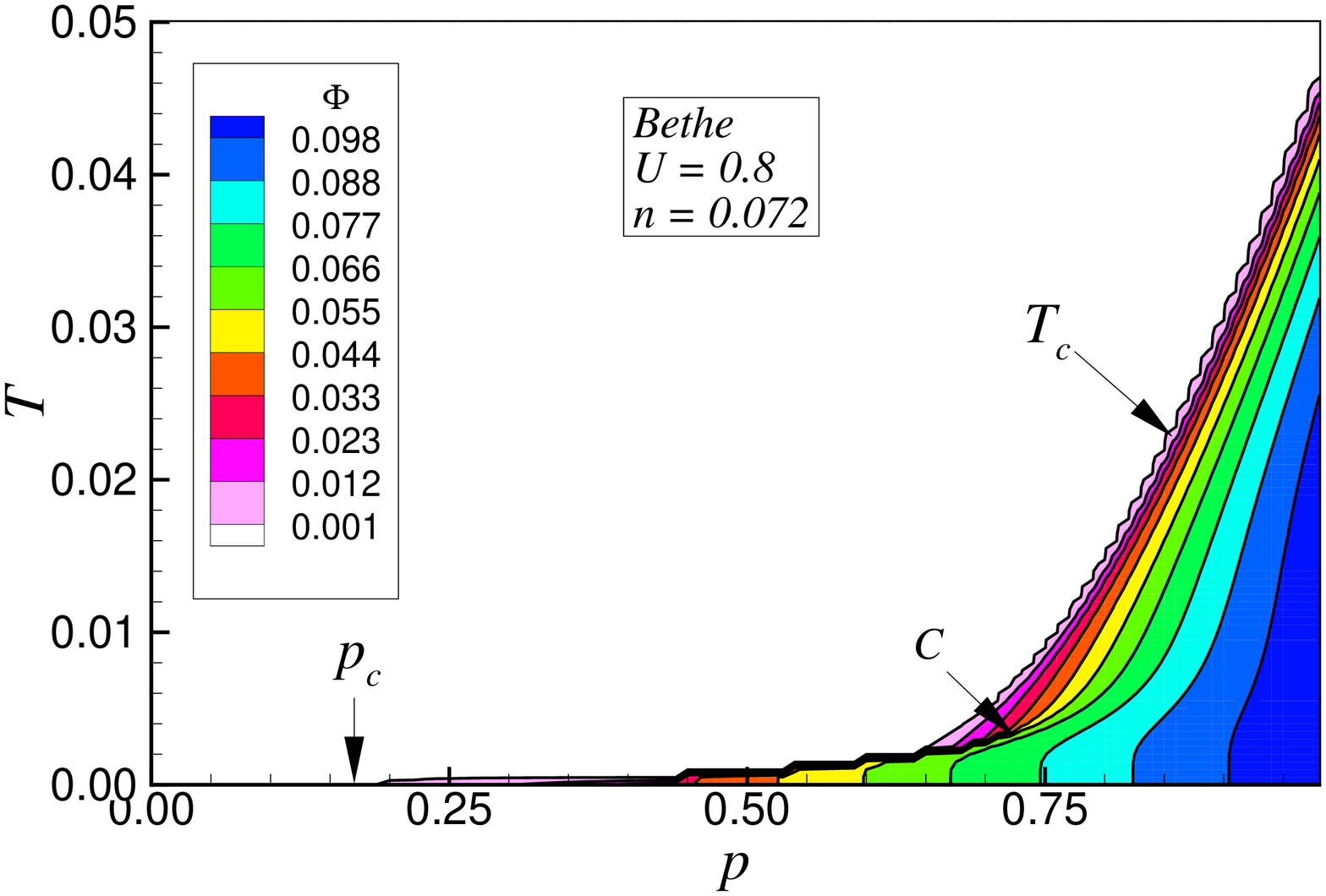}}
\subfigure[]{\epsfxsize=\figsmallwidth \epsfbox{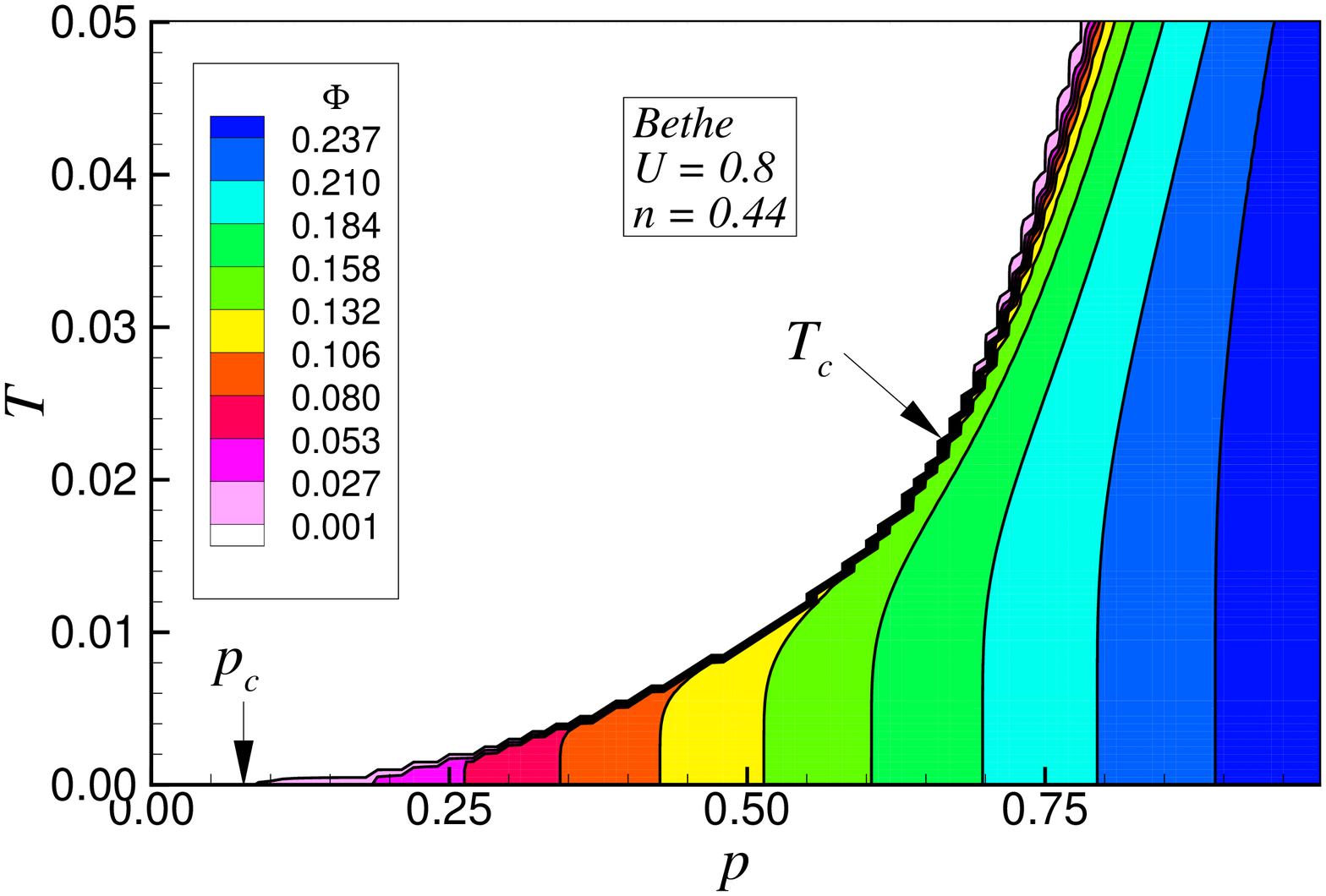}}
\subfigure[]{\epsfxsize=\figsmallwidth \epsfbox{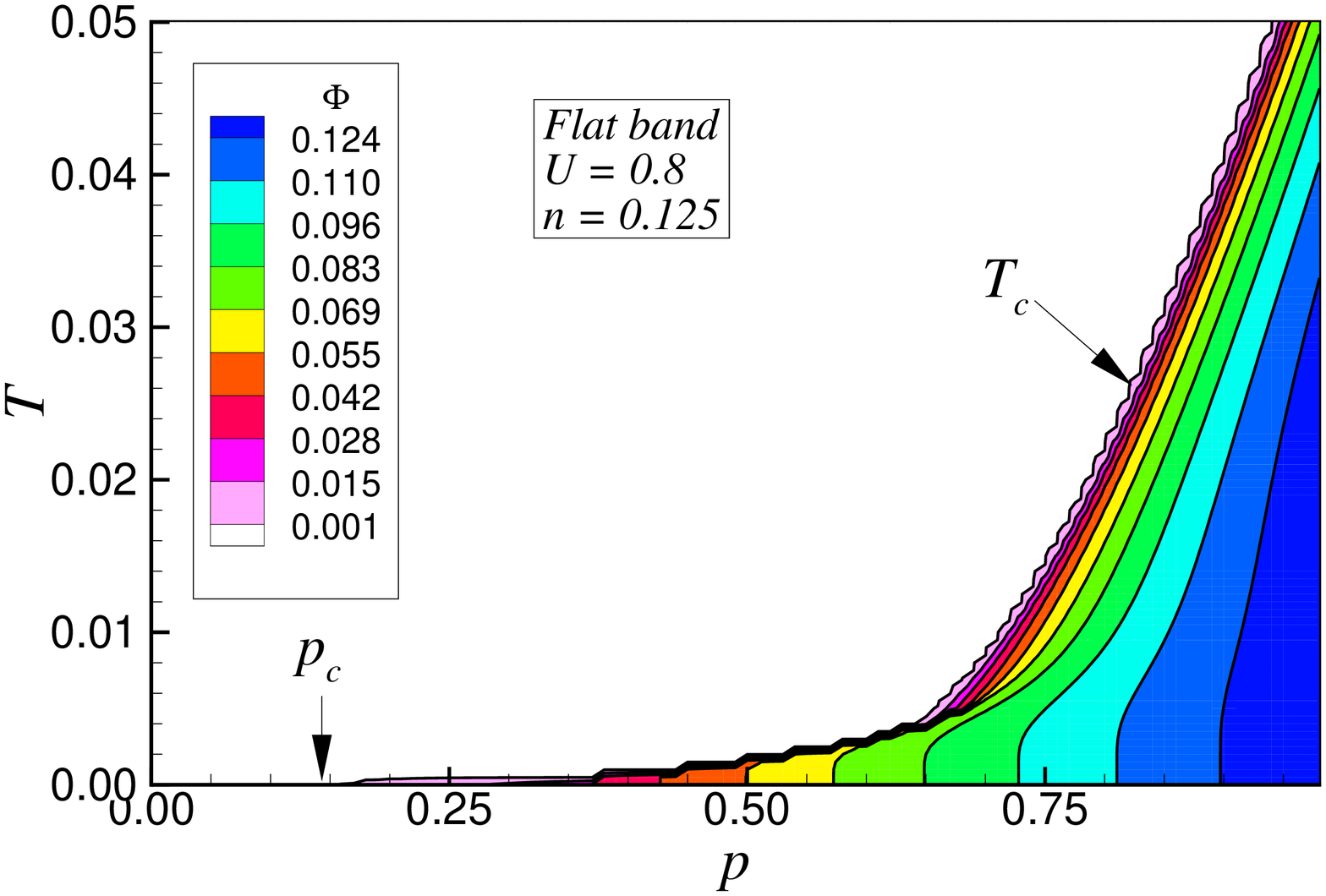}}
\subfigure[]{\epsfxsize=\figsmallwidth \epsfbox{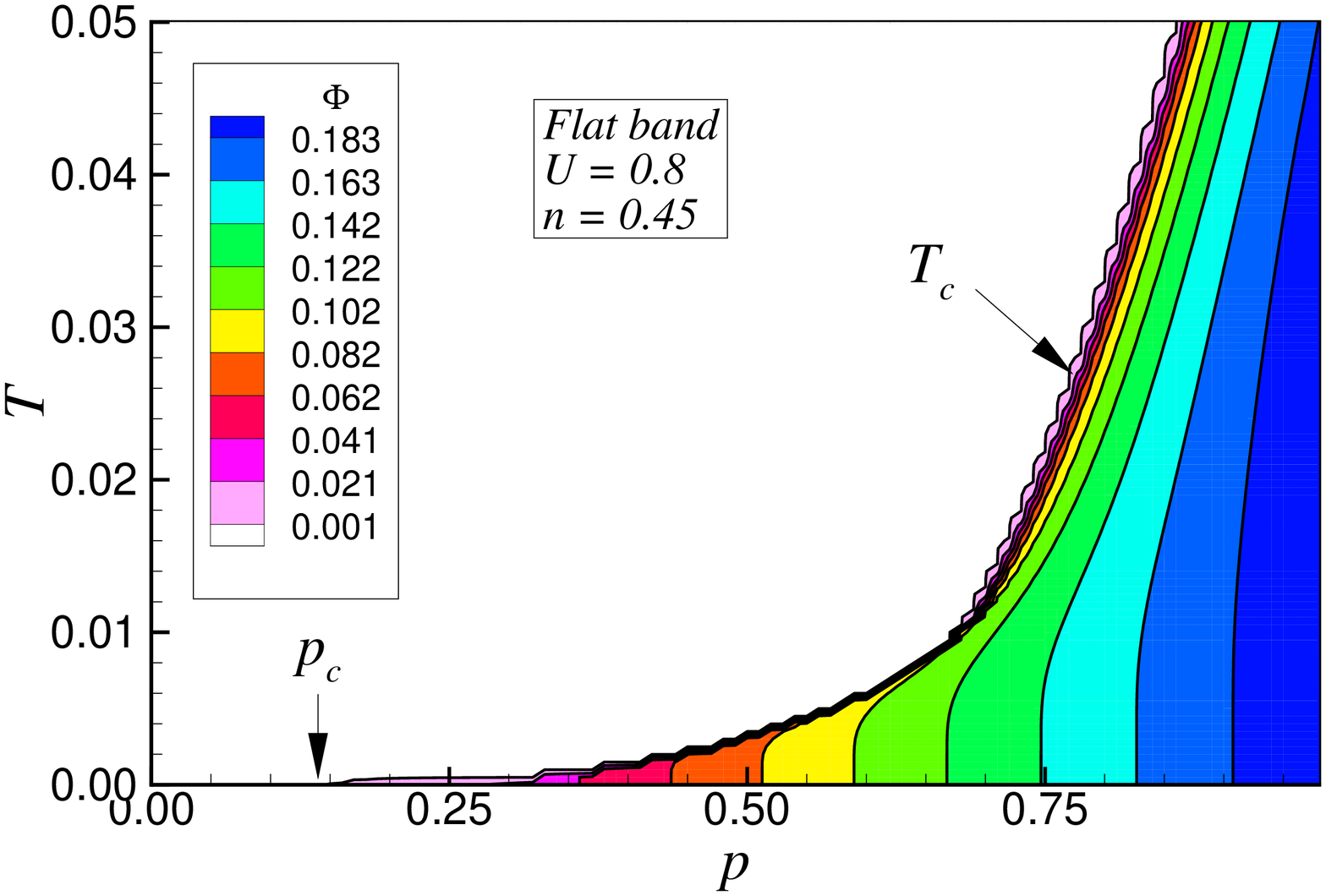}}
\caption{(color online) Phase diagram in the $p-T$ plane. Plots show contours of constant pairing  $\Phi$. The nearly straight line marked $T_c$ represents a continuous transition. The attractive interaction $U = 0.8$ in all cases. (a) Bethe lattice with $n=0.072$ (b) Bethe lattice with $n=0.44$ (c) Flat band with  $n=0.125$ (d) Flat band with $n=0.45$.}
\label{pT}
\end{figure*}

The spectral function,
\bea
N(\omega) = - \frac{1}{\pi} \Im{\mG_{11}(\omega)},
\eea
also provides interesting information regarding the nature of the
electronic state. \Fig{bspecwt} shows spectral functions for the Bethe
lattice with $U=0.8$, $n=0.44$ (results for the case of the flat band
qualitatively similar), for various values of the active fraction
$p$. The spectral function at $p=1$ has a gap with a characteristic
BCS (Bardeen-Cooper-Schrieffer) singularity\cite{Tinkham1996} in the
spectral function near the gap edges. On the other hand for $p < 1$ it
is seen that the singularity is ``smeared out'', by appearance of
``mid-gap states''. The calculation suggests a possible spatial
distribution of gaps in the system with different regions of the
lattice developing different gaps. It is, of course, not possible
within the present framework to study the gap distribution, but
further detailed simulations could throw more light on the nature of
the inhomogeneous state.

\subsection{Results at $T\ne0$}

\begin{figure}
\centerline{\epsfxsize=\figsmallwidth \epsfbox{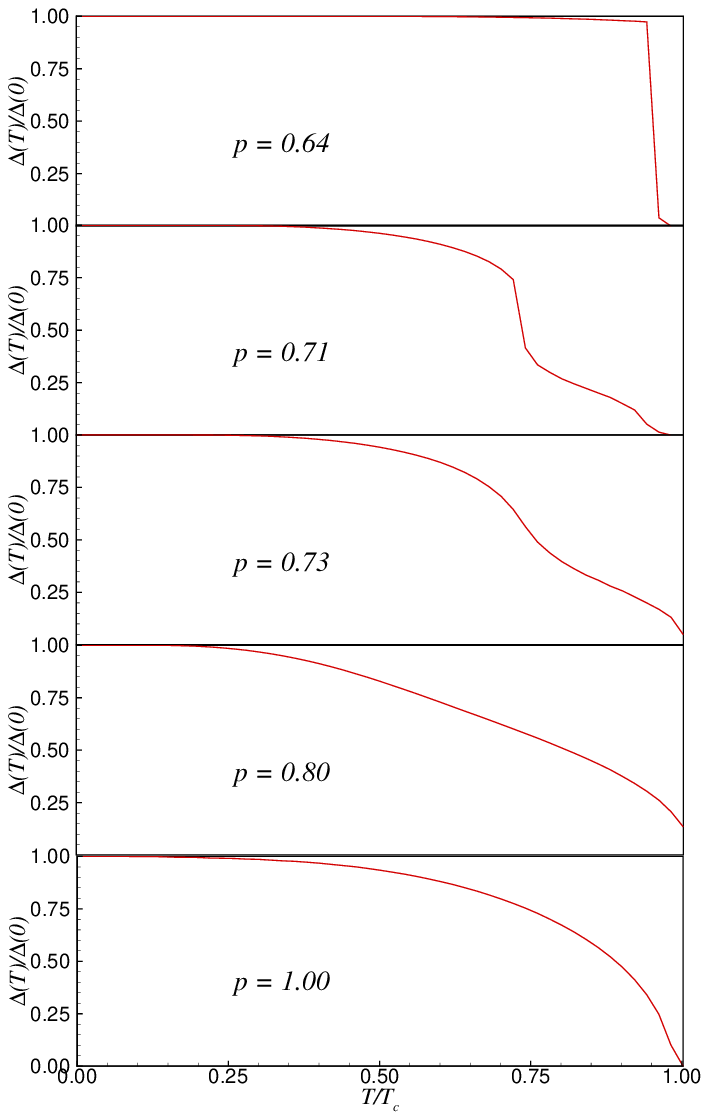}}
\caption{ (color online) Variation of $\Delta$  with temperature $T$ for the flat band $U=0.7$, $n=0.45$. The four panels show the behaviour of $\Delta$  for increasing values of the active fraction $p$.  The critical value of the active fraction for this case $p_c = 0.62$.}
\label{DeltaDepend}
\end{figure}

\begin{figure}
\centerline{\epsfxsize=\figsmallwidth \epsfbox{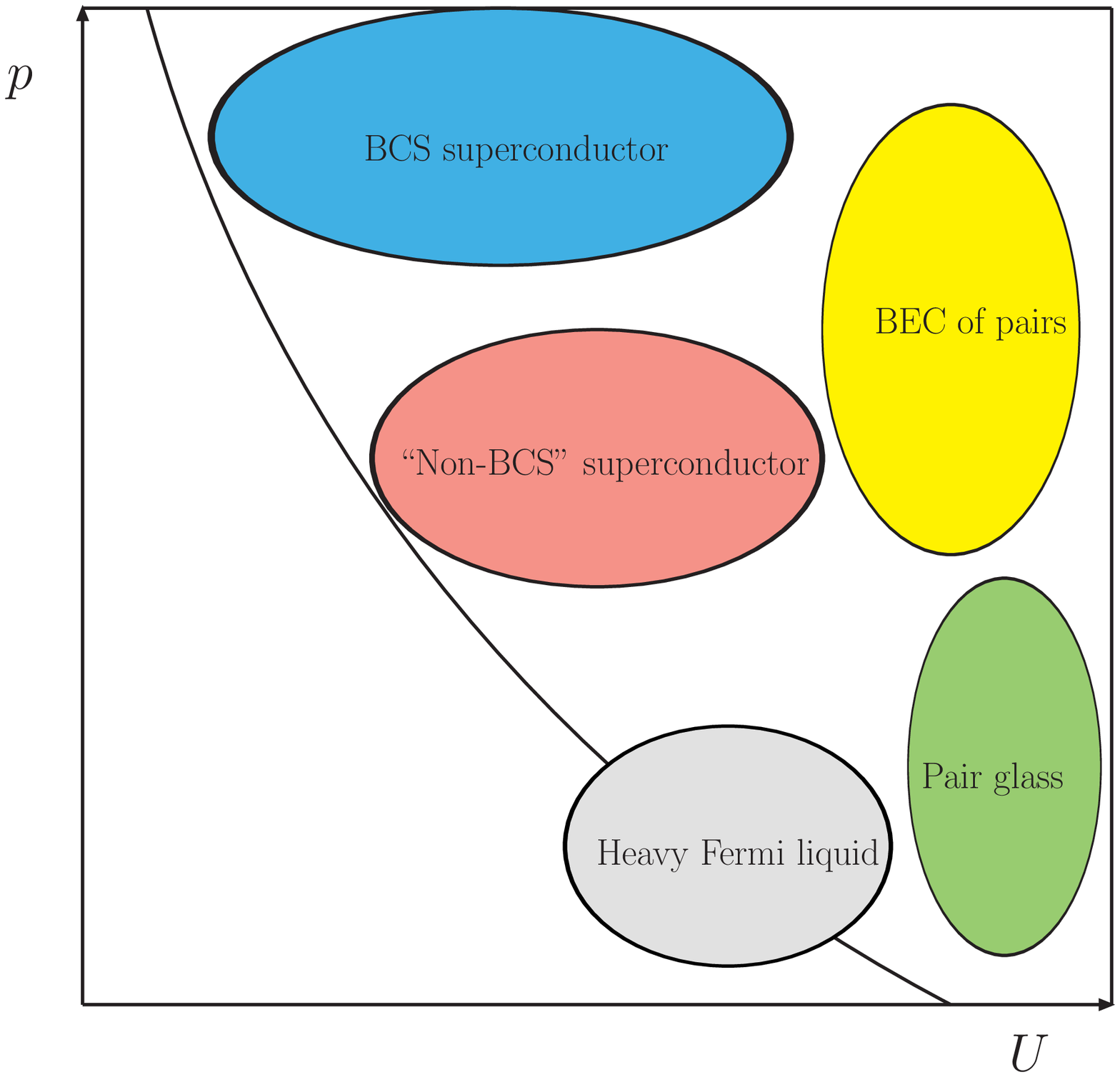}}
\caption{(color online) Schematic phase diagram of AHII. The region indicated by ``Non-BCS superconductor'' indicates the unconventional superconducting behaviour found in the calculations presented in this paper. The region indicated by ``Heavy Fermi Liquid'' can be inferred from the work of \citeasnoun{Taraphder1991}. The region with ``large $U$'' and large $p$ result in a ``BEC of pairs'' as is known from the work of \citeasnoun{Randeria1994}, and from the estimate of superfluid stiffness given in \fig{rhos}. The region with large $U$ and small $p$ is speculated to have a ``pair glass'' ground state. }
\label{SchemePD}
\end{figure}

\Fig{pT} shows the phase diagram of the AHII model in the $p-T$
plane. Several interesting features can be seen. For all active
fractions with a non-vanishing pairing amplitude at zero temperature,
there is a temperature $T_c$ at which the pairing amplitude (and
$\Delta$) vanishes. There are three regimes of active fraction that
give rise to very different finite temperature phenomenon. For active
fractions just above the critical value (marked $p_c$ in the panels in
\Fig{pT}), the transition to a regime of vanishing pairing
amplitude takes place by an abrupt (first order) transition (see, for
example, region $0.2 \le p \le 0.6$ in \Fig{pT}(a)). On increase of the
active fraction, a second regime appears (see, for example, region
$0.6 \le p \le 0.75$ in \Fig{pT}) where there are {\it two
  transitions}. In this regime, with increase of temperature from zero,
there is a first order transition where $\Phi$ (and $\Delta$)
undergoes a sudden jump and obtains a smaller {\it non zero
  value}. With further increase of temperature the pairing amplitude
vanishes continuously to zero. Interestingly, the first order line in
the $p-T$ plane appears to end at a ``critical point'' (such as that
marked by $C$ in \Fig{pT}(a)). With further increase of the active
fraction, a third regime is attained ($p > 0.75$ in \Fig{pT}(a), for
example), where $T_c$ depends essentially linearly on $p$, and the
transition is continuous.  These observations can be clearly seen by a
study of \Fig{DeltaDepend} which shows a plot $\Delta(T)/\Delta(0)$ as
a function of $T/T_c$, where the three types of behaviour are shown.
Indeed, the features are generic and do not appear to depend on the
shape of the bare density of states; they are clearly seen in both the
Bethe lattice and flat band cases. The ``sizes'' of the three regimes are,
however, strongly affected by the electron filling $n$. This is most
clearly seen in the flat band case where the critical value of $p_c$
is insensitive to electron filling. However, the finite temperature
behaviour strongly depends on electron filling -- compare \Fig{pT}(c)
and \Fig{pT}(d)), in the latter case the second regime of active
fraction with two finite temperature transitions is strongly
suppressed.

\section{Discussion and Conclusion}
\label{discussion}

This section contains a summary of the results obtained in this paper, and
a discussion of the full phase diagram of the AHII model. The present
calculation of the phase diagram is based on a single site dynamical
mean field theory like approach. The calculation shows for a certain
range of the strength of the interaction parameter $U$ which depends
on the bare band structure and electron filling, there is a critical
active fraction $p_c$ below which there is no pairing
amplitude. However, for large enough values of the interaction
parameter, even an infinitesimal value of the active fraction is
sufficient to produce a non vanishing pairing amplitude. For active
fraction $p$ greater than $p_c$ the electron spectral function shows a
features with ``mid-gap states'', and the BCS singularity of the
spectral function is smeared out. 

For cases which show a nonvanishing pairing amplitude $\Phi$, three
types of finite temperature behaviour are found. For $p$ close to
$p_c$, there is a discontinuous transition at a finite temperature,
and for $p$ close to unity, there is continuous transition (BCS like
behaviour) to a state without pairing amplitude. There is a
intermediate range of active fractions, where the transition to a non
paired state takes place in two steps -- ``non-BCS behaviour''. As the
temperature is increased, there is a first order transition to a state
with smaller $\Phi$. Further increase of temperature causes a
continuous transition to a state with no pairing amplitude. It is
tempting to speculate that the state attained up on the first order
transition has a pairing amplitude, but no superconductivity. The
physical picture of such a state is that of ``puddles of electrons''
with non-zero pairing amplitude without a global phase necessary for
superconductivity. Such a state is likely to show ``psuedo-gap'' like
features, for example, a reduced spin susceptibility. Clearly, this
finding of the present calculation needs more attention, and the
region of the phase diagram where this phenomenon is found needs
further detailed investigation. It is interesting to note that
calculations based on BdGMF\cite{Aryanpour2007} also show a regime of
$U$ and $p$ which show anomalous behaviour of $\Delta$ as a function
of $T$.

The present formulation is based on a single site formulation and
averages over all the spatial correlations. However, as noted above, a
very interesting region in the phase diagram is revealed, and suggests
possibility for further investigation. Further, the approximate
treatment based on the saddle point approximation does not include
quantum fluctuations. It is believed that the inclusion of these
quantum fluctuation effects are not likely to change the qualitative
features of the present single site calculation; this is suggested by
the iterated perturbation theory based dynamical mean field theory of
the attractive Hubbard model.\cite{Garg2005} It must be noted that
most of the previous work cited above are based on two dimensional
systems, mostly square lattices. Long wavelength fluctuations, crucial
in two dimensional systems, cannot be treated within the present
framework. The present work, therefore, is more applicable to higher
dimensional systems such as negative-$U$ centre systems etc.

\begin{figure}
\centerline{\epsfxsize=\figsmallwidth \epsfbox{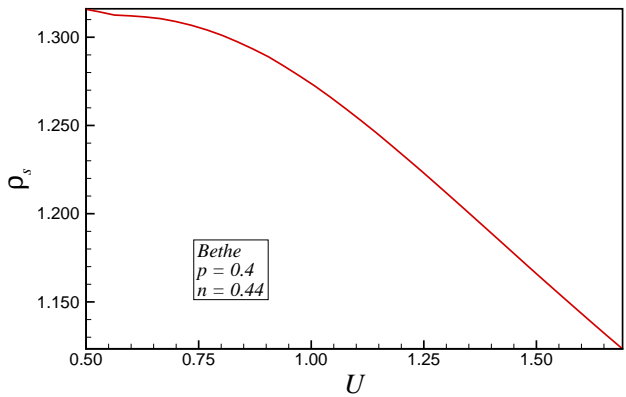}}
\caption{ (color online) Dependence of the superfluid density $\rho_s$ on the interaction strength $U$ in the Bethe lattice with $n=0.44$ and $p=0.4$.}
\label{rhos}
\end{figure}

The paper is concluded with a discussion of the complete phase diagram
of the AHII model. Based on the calculations presented here and on
published results quoted earlier, the nature of the electronic state in
different regions of the parameter states can be inferred.  In the
regions with nonzero pairing amplitude (as obtained from the present
calculation), different types of electronic states may be found as
indicated in \Fig{SchemePD}. For intermediate $U$ (compared to the
bare bandwidth) and large $p$ a BCS superconductor is obtained. On the
other hand for smaller value of the active faction $p$, a ``non-BCS''
superconductor is obtained (as discussed above), with the possibility
of a high temperature ``pseudo-gap'' phase. For larger values of the
interaction strength there is a crossover from BCS to BEC like
behaviour where the electrons form pairs and Bose
condense.\cite{Randeria1994} In the present calculation, this
behaviour is inferred by the calculation of an estimate (based on the
kinetic energy) of the superfluid density.\cite{Garg2005} As shown in
\Fig{rhos}, the superfluid density $\rho_s$ falls with increasing $U$
indicating a crossover from BCS to BEC (Bose-Einstein condensation)
behaviour. The effect of the random attractive interaction in this BEC
regime needs a more careful investigation than that given here. For
low values of the active fraction and intermediate values of the
interaction strength, the most likely ground state is a heavy Fermi
liquid\cite{Taraphder1991} engendered by the charge Kondo effect. It
is also possible that for small fillings, large interaction strengths,
one could obtain a pair glass, where electrons are localized at the
negative $U$ centers. Clearly, the ``boundaries'' of the regions
indicated in the phase diagram will be determined by the third factor
in the problem, namely electron filling. The nature of the electronic
states in different parameter regimes of the AHII model does resemble
various systems discussed in the introductory section. It will be
interesting also to explore the possibility of a direct experimental
realization of the AHII model in cold atom optical
lattices.\cite{Georges2007,Bloch2007,Ketterle2008}

\section*{Acknowledgement}

The author thanks A.~Ghosh,  H.~R.~Krishnamurthy, T.~V.~Ramakrishanan and T.~Senthil for discussions, and  S.~Bhowmick and S.~Pathak for comments on the manuscript. Generous support for this work by DST, India through a Ramanujan grant is gratefully acknowledged.

\bibliography{ahub}

\end{document}